\documentclass[twocolumn,showpacs,preprintnumbers,amsmath,amssymb]{revtex4}
\usepackage{color}
\usepackage{bm, graphicx, amsmath}
\usepackage{hyperref}
\begin{document}

\title{Bell inequalities from group actions: Three parties and non-Abelian groups}
\author{V. U\u{g}ur G\"{u}ney and Mark Hillery}
\affiliation{Department of Physics, Hunter College of the City University of New York, 695 Park Avenue, New York, NY 10065 USA \\ Graduate Center of the City University of New York, 365 Fifth Avenue, New York, NY 10016}

\begin{abstract}
In a previous publication, we showed how group actions can be used to generate Bell inequalities.  The group action yields a set of measurement probabilities whose sum is the basic element in the inequality.  The sum has an upper bound if the probabilities are a result of a local, realistic theory, but this bound can be violated if the probabilities come from quantum mechanics.  In our first paper, we considered the case of only two parties making the measurements and single-generator groups.  Here we show that the method can be extended to three parties, and it can also be extended to non-Abelian groups.  We discuss the resulting inequalities in terms of nonlocal games.
\end{abstract}


\maketitle

\section{Introduction}
A Bell inequality is an inequality containing probabilities of measurement results that will be obeyed by probabilities resulting from a local, realistic theory \cite{bell}.  Initial interest in them was confined to people working in the foundations of quantum mechanics, but more recently they have provided the basis for protocols in quantum cryptography and for tests of entanglement.  There is now an extensive literature on the subject, and considerable progress has been made in classifying and tabulating Bell inequalities.  Two recent reviews provide a good overview of the field \cite{brunner,liang}.

The standard scenario for a Bell inequality is that there are $N$ parties making measurements, each party can make one of $M$ possible measurements, and each measurement has $K$ outcomes.  Perhaps the most famous Bell inequality, the Clauser-Horne-Shimony-Holt (CHSH) inequality, is for the case $N=M=K=2$ \cite{clauser}.  Kaszlikowsi, et al.\ showed that by increasing the number of outcomes, $K$, one could more strongly violate local realism \cite{kaszlikowski}. The case of full correlation Bell inequalities for $M=K=2$ and general $N$ has been fully characterized by Werner and Wolf \cite{werner}.  Bell inequalities for the case $N=2$, $M=2$, and general $K$ were found by Collins, et al.\ \cite{collins}, and this was generalized to the case of general $N$, $M=2$, and general $K$ by Son, et.\ al. \cite{son}.

Recently, an approach to Bell inequalities based on graph theory was developed by Cabello, Severini, and Winter \cite{cabello}.  The vertices of a graph correspond to events, where an event is a particular choice of measurement and a measurement outcome for each party.  The probabilities of these events are what appear in the inequality, in particular their sum.  Two vertices are connected by an edge if the events corresponding to them cannot be true simultaneously.  The properties of the graph can be used to find an upper bound to the classical sum of the probabilities, that is the sum when the probabilities come from a local, realistic theory,  and also an upper bound to the quantum sum, where the probabilities come from quantum mechanics.

In a previous paper, we explored an approach to Bell inequalities based on group actions of single generator Abelian groups \cite{guney}.  In that case the Bell inequalities also involve sums of the probabilities of events, but instead of starting from a graph, we start from a group.  The events are generated by the application of operators that form a representation of a group to an initial state.  As an example, suppose we have two parties, Alice and Bob, and they each have a qubit so that their joint states are elements of a tensor product Hilbert space, $\mathbb{C}^{2}\otimes \mathbb{C}^{2}$.  We will denote the computational, or $z$, basis of $\mathbb{C}^{2}$ by $\{ |0\rangle , |1\rangle \}$ and the $x$ basis by $\{ |\pm x\rangle = (|0\rangle \pm |1\rangle )/\sqrt{2} \}$.  Consider the operator $U= |+x\rangle\langle +x| -i|-x\rangle\langle -x|$, and note that $U^{2}= |+x\rangle\langle +x|-|-x\rangle\langle-x| = \sigma_{x}$, and $U^{4}=I$.  We have that $\{j\rightarrow U^{j}|j=0,1,2,3\}$ is a representation of the cyclic group $\mathbb{Z}_{4}$, the group of addition modulo $4$, and so is $\{ j\rightarrow U^{j}\otimes U^{j}|j=0,1,2,3\}$.  The map from $|\Psi\rangle \in \mathbb{C}^{2}\otimes \mathbb{C}^{2}$ and $j\in \mathbb{Z}_{4}$ to the state $U^{j}\otimes U^{j} |\Psi\rangle$ is an example of a group action \cite{rotman}.

The definition of a group action is the following.  If $G$ is a group and $X$ is a set, a group action is a function $\alpha: G\times X\rightarrow X$ such that $\alpha (e,x)=x$ and $\alpha (g,\alpha (h,x))=\alpha (gh,x)$.  Here $e,g,h\in G$ and $e$ is the identity element of the group.  The subset of $X$ given by $\{ \alpha (g,x)| g\in G\}$ is called the orbit of $x$.  Any two orbits are either distinct or identical, so the set of orbits forms a partition of $X$.  In the case of our example, the orbits will be sets of the form $\{ U^{j}\otimes U^{j} |\Psi\rangle |j=0,1,2,3\}$.  Now let us set $|\Psi\rangle = |0,0\rangle$, in which case the orbit is $\{ |0,0\rangle ,  |v_{0},v_{0}\rangle , |1,1\rangle , |v_{1},v_{1}\rangle \}$, where $|v_{j}\rangle =U|j\rangle$ for $j=0,1$.  Define the observables for Alice to be $a_{0}=|1\rangle\langle 1|$ and $a_{1}=|v_{1}\rangle\langle v_{1}|$ and similarly for Bob, $b_{0}=|1\rangle\langle 1|$ and $b_{1}=|v_{1}\rangle\langle v_{1}|$.  The orbit then corresponds to the events $(a_{0}=0, b_{0}=0)$, $(a_{1}=0,b_{1}=0)$, $(a_{0}=1, b_{0}=1)$, and $(a_{1}=1,b_{1}=1)$.  We can also choose a second orbit starting with the state $|0,v_{0}\rangle$, which generates four more events.  We will not demonstrate this here (a closely related example appeared in \cite{guney}), but the sum of the probabilities for these eight events cannot be larger than $3$ if the probabilities come from a local realistic theory, whereas it can reach the value of $2+\sqrt{2}$ if the probabilities come from quantum mechanics.

In this paper we would like to extend these results in two directions.  First, we previously only considered the case of two parties.  In the next section we again consider cyclic groups but for the case of three parties.  In the following section, we go back to the case of two parties, but look at non-Abelian groups, in particular dihedral groups.  We compare the nonlocal games that result from Bell inequalities for the cyclic group $\mathbb{Z}_{6}$ and the dihedral group $D_{3}$, both of which have $6$ members.  Finally, we show how group representation theory can be used to find quantum states that violate a Bell inequality that results from the dihedral group $D_{6}$, a group with $12$ members.  

\section{Three-party case}
We shall consider a scenario in which three parties, Alice, Bob, and Charlie, share a system of three particles, each party possessing one of the particles.  Each of them can measure one of two observables, and for each observable the possible values for the result of the measurement are $0$, $1$, or $2$.  Alice's observables are $a_{0}$ and $a_{1}$, Bob's are $b_{0}$ and $b_{1}$, and Charlie's are $c_{0}$ and $c_{1}$.

In the quantum mechanical version of this scenario, Alice, Bob, and Charlie, each has a qutrit.  The computational basis is $\{ |j\rangle\, |\, j=0,1,2\}$ and corresponds to the observables $a_{0}$, $b_{0}$, and $c_{0}$; for example, $a_{0}=|1\rangle\langle 1| + 2|2\rangle\langle 2|$, and similarly
for $b_{0}$ and $c_{0}$.  In order to define a second basis, consider the operator
\begin{equation}
U=|w_{0}\rangle\langle w_{0}| + e^{-i\pi /3} |w_{1}\rangle\langle w_{1}| + e^{i\pi /3}|w_{2}\rangle\langle w_{2}| ,
\end{equation}
where 
\begin{equation}
|w_{j}\rangle = \frac{1}{\sqrt{3}} \sum_{k=0}^{2} e^{2\pi ijk/3} |k\rangle ,
\end{equation}
for $j=0,1,2$.  We have that $U^{2}=T$, where $T|j\rangle = |j+1\rangle$, and the addition is modulo $3$.  Note that $U^{6}=I$, which implies that $\{ U^{m}|\, m=0,1,\ldots 5\}$ is a representation of the group $\mathbb{Z}_{6}$, addition modulo $6$.  We now define a second basis, $\{ |v_{j}\rangle = U|j\rangle\, |\, j=0,1,2\}$, and this basis corresponds to the observables $a_{1}$, $b_{1}$, and $c_{1}$, where, for example, $a_{1}=|v_{1}\rangle\langle v_{1}| + 2|v_{2}\rangle\langle v_{2}|$.  Each of the three parties can measure their qutrit in either of the two bases.  We now choose four states, and apply to each of them the operator $(U\otimes U\otimes U)^{m}$, for $m=0, 1, \ldots 5$, which generates a total of $24$ states.  The four states are:
\begin{equation}
\label{3-party-orbits}
|021\rangle \hspace{5mm} |00v_{1}\rangle \hspace{5mm} |0v_{0}0\rangle \hspace{5mm} |v_{0}20\rangle .
\end{equation}
Note that in the first state all three bases are the same, in the second, the first two bases are the same, in the third, the first and third bases are the same, and in the fourth the second and third bases are the same.  This feature remains the same under application of $(U\otimes U\otimes U)^{m}$, so, for example, for all states generated by application of this operator to the first state the bases will be the same.  Note that these basis combinations exhaust all possible choices of measurement bases by the parties.  Because there are only two measurement bases, either each party chooses the same basis or two of them do and the third party chooses a different one.  This second alternative can happen in three different ways.  This results in $24$ three-qutrit states each of which is a product of single-qutrit states from one of the two bases. 

Note that $m\in \mathbb{Z}_{6}\rightarrow (U\otimes U\otimes U)^{m}$ is also a representation of $\mathbb{Z}_{6}$.  Application of these operators to $\mathbb{C}^{3}\otimes \mathbb{C}^{3}\otimes \mathbb{C}^{3}$ defines a group action, and the set resulting from the application of all six operators to a particular state defines the orbit associated with that state.  Two orbits are either distinct or identical.  Each of the four states in Eq.\ (\ref{3-party-orbits}) generates an orbit.  These states were found by means of a random search in the set of states that are threefold tensor products of the states in the computational and $v$ bases.  The search identified sets of states whose orbits lead to Bell inequality violations.  More details of how the random searches in this paper are performed can be found in Appendix A.

Each of the $24$ states in our set corresponds to a particular choice of measurements by the three parties and a particular set of measurement results.  For example, the state $|00v_{1}\rangle$ corresponds to Alice measuring $a_{0}$ and obtaining $0$, Bob measuring $b_{0}$ and obtaining $0$, and Charlie measuring $c_{1}$ and obtaining $1$.  In order to maximize the sum of probabilities corresponding to these measurement choices and the specified results, we need to find the state, $|\phi\rangle$, that maximizes the expectation value $\langle\phi |A| \phi\rangle$, where
\begin{equation}
A= \sum_{m=0}^{5} (U\otimes U\otimes U)^{m} L (U^{\dagger}\otimes U^{\dagger}\otimes U^{\dagger})^{m} ,
\end{equation}
and
\begin{eqnarray}
L & = & |021\rangle\langle 021|+  |00v_{1}\rangle\langle 00v_{1}| + |0v_{0}0\rangle\langle 0v_{0}0| 
\nonumber \\
& & + |v_{0}20\rangle\langle v_{0}20| .
\end{eqnarray}
The expectation value of $A$ in the state $|\phi\rangle$ is just the sum of the 24 probabilities if Alice, Bob, and Charlie share the three-qutrit state $|\phi\rangle$.  The largest value of the expectation value occurs when $|\phi\rangle$ is the eigenstate of $A$ corresponding to its largest eigenvalue. 

We want to find the largest eigenvalue of $A$, but before proceeding let us note something that will simplify the calculation.  If we define $B=U\otimes U\otimes U$ then we see that the eigenstates of $B$ are states of the form $|w_{j}\rangle |w_{k}\rangle |w_{l}\rangle$, and the possible eigenvalues are $1$, $-1$, $e^{\pm i\pi /3}$, and $e^{\pm 2\pi i/3}$, all of which are degenerate.  Let $P_{\lambda}$ be the projection onto the subspace corresponding to the eigenvalue, $\lambda$, of $B$.  Because $[P_{\lambda},B]=0$, we have that
\begin{eqnarray}
\label{3-party-proj}
A & = & \left(\sum_{\lambda}P_{\lambda}\right) \sum_{j=0}^{5} B^{j}L(B^{\dagger})^{j}\left( \sum_{\lambda^{\prime}} P_{\lambda^{\prime}} \right) \nonumber \\
 & = & \sum_{\lambda} \sum_{\lambda^{\prime}} \left( \sum_{j=0}^{5} \lambda^{j}(\lambda^{\prime \ast})^{j} \right) P_{\lambda}L P_{\lambda^{\prime}} \nonumber \\
 & = & \sum_{\lambda} \sum_{\lambda^{\prime}} 6 \delta_{\lambda , \lambda^{\prime}} P_{\lambda}L P_{\lambda^{\prime}} \nonumber \\
 & = & 6 \sum_{\lambda} P_{\lambda}L P_{\lambda} .
\end{eqnarray}
Therefore, in order to diagonalize $A$ we only have to diagonalize it within the subspaces corresponding to the eigenvalues of $B$. 

We find that the eigenvector corresponding to the maximum eigenvalue of $A$ lies in the subspace where $B$ has an eigenvalue of $1$.  This space is seven dimensional, so we are faced with diagonalizing a seven-dimensional matrix.  However, because of the form of the matrix,  
\begin{equation}
M=\sum_{j=1}^{4} |\mu_{j}\rangle\langle \mu_{j}| ,
\end{equation}
where $|\mu_{1}\rangle = P_{1}|021\rangle$, $|\mu_{2}\rangle = P_{1}|00v_{1}\rangle$, $|\mu_{3}\rangle = P_{1}|0v_{0}0\rangle$, and $|\mu_{4}\rangle = P_{1}|v_{0}20\rangle$, the problem can be reduced to a four-dimensional one.  If we express the eigenvector as $\sum_{j=1}^{4} c_{j}|\mu_{j}\rangle$, then the eigenvalue equation becomes
\begin{equation}
\sum_{k=1}^{4} |\mu_{k}\rangle \left( \sum_{j=1}^{4}c_{j}\langle \mu_{k}|\mu_{j}\rangle \right) = \lambda
\sum_{k=1}^{4}c_{k}|\mu_{k}\rangle .
\end{equation}
Finding the overlaps of the vectors, we obtain
\begin{equation}
\frac{1}{27}\left(\begin{array}{cccc} 7 & 2 & -1 & -1 \\ 2 & 7 & -1 & -1\\ -1 & -1 & 7 & -1 \\ -1 & -1 & -1 & 7 \end{array} \right) \left( \begin{array}{c} c_{1} \\ c_{2} \\ c_{3} \\ c_{4} \end{array} \right) = \lambda \left( \begin{array}{c} c_{1} \\ c_{2} \\ c_{3} \\ c_{4} \end{array} \right) .
\end{equation}
The largest eigenvalue is $10/27$, which gives $6(10/27) = 20/9$ (see Eq.\ (\ref{3-party-proj})) as the largest eigenvalue of $A$. The eigenvector is given in the computational basis by
\begin{eqnarray}
\label{3-party-state}
|\phi\rangle & = & \frac{1}{30\sqrt{3}} [ -10 (|000\rangle + |111\rangle + |222\rangle ) + 14 ( |001\rangle
\nonumber \\ 
& & + |112\rangle +  |220\rangle ) + 11(|002\rangle + |110\rangle \nonumber \\
& &  + |221\rangle ) -7 ( |010\rangle + |121\rangle + |202\rangle ) \nonumber \\
& & - ( |011\rangle + |022\rangle + |100\rangle + |122\rangle  \nonumber \\
&& +|200\rangle + |211\rangle ) -4(|012\rangle + |020\rangle \nonumber \\
& & + |101\rangle + |120\rangle + |201\rangle + |212\rangle )\nonumber \\
& & + 20 (|021\rangle + |102\rangle + |210\rangle ) ] .
\end{eqnarray}
The maximum value of the sum of the classical probabilities is $2$, so we get a violation.

The classical bound is found by assuming that there is a joint distribution for all of the variables,$P(a_{0},b_{0},c_{0};a_{1},b_{1},c_{1})$.  Then each of the $24$ measurement probabilities can be expressed in terms of joint distribution, and their sum can be expressed as 
\begin{equation}
\sum_{j=0}^{1} \sum_{a_{j},b_{j},c_{j}=0}^{2} c_{a_{0},b_{0},c_{0};a_{1},b_{1},c_{1}} P(a_{0},b_{0},c_{0};a_{1},b_{1},c_{1}) ,
\end{equation}
where the $c_{a_{0},b_{0},c_{0};a_{1},b_{1},c_{1}}$ are integers.  The sum is maximized when the probability distribution is chosen to have the value $1$ for one of the values of $(a_{0},b_{0},c_{0},a_{1},b_{1},c_{1})$ corresponding to the largest value of $c_{a_{0},b_{0},c_{0}:a_{1},b_{1},c_{1}}$, which implies that the maximum value of the sum is just the maximum value of $c_{a_{0},b_{0},c_{0}:a_{1},b_{1},c_{1}}$.  In this case we find that the largest value is $2$, which is, therefore, the largest value the sum of the probabilities can assume.

It is also possible to phrase this inequality as a nonlocal game.  Alice, Bob, and Charlie each receive a bit, $s$, $t$, and $u$, respectively, where $s,t,u=0$ or $1$ from an arbitrator.  Each then transmits to the arbitrator, a $0$, a $1$, or a $2$.  The arbitrator then decides whether they have won the game.  The wining conditions depend on both $s$, $t$, and $u$ and the values returned by Alice, Bob and Charlie.  For this game, the winning conditions are listed in Table 1.  The values of $(s,t,u)$ are listed on the left, and the corresponding winning combinations of the values sent by Alice, Bob and Charlie are listed on the right.  The values of $(s,t,u)$ are grouped according to which of the values are the same.  The first two have all the values the same, the next two have their first two values the same, the next two have the first and third values the same, and the last two have the second and third values the same.  The sum of the digits in the winning sequences modulo $3$ in each row are the same, $0$ in rows $1,\ 2,\ 5,\ 8$, $1$ in rows $3,\ 6$, and $2$ in rows $4,7$.
\newline

\begin{table}
\centering
\begin{tabular}{|c|c|} \hline
s,t,u & Alice, Bob, Charlie \\ \hline
$000$ & $021$, $102$, $210$,  \\ $111$ & $021$, $102$, $210$ \\ $001$ & $001$, $112$, $220$ \\
$110$ & $002$, $110$, $221$ \\ $010$ & $000$, $111$, $222$ \\ $101$ & $010$, $121$, $202$ \\
$100$ & $020$, $101$, $212$ \\ $011$ & $012$, $120$, $201$ \\ \hline
\end{tabular} 
\caption{Winning conditions for the nonlocal game}
\end{table}  

Let us now look at classical strategies for winning this game.  We assume that all of the eight possible values of the triplet $(s,t,u)$ are equally likely.  We then note that if Alice always returns $0$, Bob always returns $2$ and Charlie always returns $1$, then they win the game with a probability of $1/4$. The next step is to show that this is the best that can be done.  A deterministic classical strategy can be specified by three functions, $f_{A}(s)$, $f_{B}(t)$, and $f_{C}(u)$.  Each of these functions takes values in the set $\{ 0,1,2\}$, and if Alice receives the value $s$, she returns $f_{A}(s)$, and similarly for Bob and Charlie.  Let $F(a,b,c;s,t,u)$, where $a,b,c\in \{ 0,1,2\}$ be equal to $1$ when $(a,b,c;s,t,u)$ is a winning condition for the game and $0$ otherwise.  Then the success probability for the strategy represented by $f_{A}(s)$, $f_{B}(t)$, and $f_{C}(u)$ is
\begin{equation}
\label{classwinstrat}
\frac{1}{8} \sum_{a,b,c=0}^{2} \sum_{s,t,u=0}^{1} F(a,b,c;s,t,u) \delta_{a,f_{A}(s)} \delta_{b,f_{B}(t)} \delta_{c,f_{C}(u)} .
\end{equation}
Let us compare this with the expression for the sum of our $24$ probabilities, which can be expressed as
\begin{equation}
\sum_{a,b,c=0}^{2} \sum_{s,t,u=0}^{1} F(a,b,c;s,t,u) p(a_{s}=a,b_{t}=b,c_{u}=c) ,
\end{equation}
which, we found is bounded above by $2$ if the probabilities $p(a_{s}=a,b_{t}=b,c_{u}=c)$ are derived from a joint distribution.  Noting that $\delta_{a,f_{A}(s)} \delta_{b,f_{B}(t)} \delta_{c,f_{C}(u)}$ can be derived from a joint distribution, in particular
\begin{eqnarray}
P(a_{0},b_{0},c_{0};a_{1},b_{1},c_{1}) & = & \delta_{a_{0},f_{A}(0)} \delta_{a_{1},f_{A}(1)} \nonumber \\ & & \delta_{b_{0},f_{B}(0)}  \delta_{b_{1},f_{B}(1)} \nonumber \\
& & \delta_{c_{0},f_{C}(0)}\delta_{c_{1},f_{C}(1)} ,
\end{eqnarray}
we see that the sum in Eq.\ (\ref{classwinstrat}) is less than or equal to $2$, which implies that the classical winning strategy must be less than or equal to $1/4$.  A deterministic strategy is an optimal one \cite{cleve}, so this means that the maximum classical probability of winning the game is $1/4$.

In the quantum strategy, Alice, Bob, and Charlie share the quantum state in Eq.\ (\ref{3-party-state}) and make measurements on their respective quitrits.  Which measurement they make is dictated by the value of the bit they receive from the arbitrator.  In particular, Alice measures $a_{s}$, Bob measures $b_{t}$, and Charlie measures $c_{u}$.  They then just send the results of their measurements to the arbitrator.  Their probability of winning is just $1/8$ times the sum of the probabilities of the winning configurations, which we have seen is $20/9$.  This gives an overall probability of $5/18$, which is approximately $0.28$, and this is greater than the winning probability of the classical strategy. 

\section{Dihedral groups}
So far, in this paper and in our previous one, we have only made use of abelian groups to generate Bell inequalities.  Now we would like to show, by way of an example, that non-abelian groups can also be used.  We shall give two examples using dihedral groups.  There is a family of dihedral groups, and the dihedral group $D_{n}$ is the group consisting of rotations and reflections in the plane that leave an $n$-sided regular polygon invariant.  It is generated by two elements,  a rotation, $r$, by angle $2\pi /n$ and a reflection $s$.  For example, in the case of an equilateral triangle, the reflection would be about an axis passing through one of the vertices and the midpoint of the opposite side.  $D_{n}$ has $2n$ elements, $e$, $r^{j}$ for $j=1,2,\ldots n-1$, and $r^{j}s$ for $j=1,2,\ldots n-1$, where $e$ is the identity element.  The group is specified by its presentation, which consists of the elements $r$ and $s$, and the relations $r^{n}=e$, $s^{2}=e$, and $srs=r^{-1}$.

Our approach will make use of the representations of dihedral groups.  We will look at two examples, one making use of $D_{3}$ and the other making use of $D_{6}$.

\subsection{$D_{3}$}
The group $D_{3}$ consists of rotations and reflections in the plane that leave an equilateral triangle invariant.  It consists of the elements $\{ e,r,r^{2}, s, rs, r^{2}s \}$, where $r^{3}=e$ and $s^{2}=e$.  The group has three conjugacy classes $C_{e}=\{ e\}$, $C_{r}=\{ r,r^{2}\}$, and $C_{s}=\{ s, rs, r^{2}s \}$.  It has three irreducible representations, $\Gamma^{(j)}$ for $j=1,2,3$, where $\Gamma^{(1)}$ and $\Gamma^{(2)}$ are one-dimensional and $\Gamma^{(3)}$ is two dimensional.  The character table for the group is given in Table 2.  

\begin{table}
\centering
\begin{tabular}{|c|c|c|c|} \hline
 & $C_{e}$ & $C_{r}$ & $C_{s}$ \\ \hline $\Gamma^{(1)}$ & $1$ & $1$& $1$ \\ \hline $\Gamma^{(2)}$ & $1$ & $1$ & $-1$ \\ \hline $\Gamma^{(3)}$ & $2$ & $-1$ & $0$ \\ \hline 
\end{tabular}
\caption{Character table for $D_{3}$.}
\end{table}

For the representation $\Gamma^{(3)}$, we can take for the matrices corresponding to $r$ and $s$ 
\begin{equation}
U=\left( \begin{array}{cc} -1/2 & -\sqrt{3}/2 \\ \sqrt{3}/2 & -1/2 \end{array} \right) \hspace{5mm} V=\left( \begin{array}{cc} 1 & 0 \\ 0 & -1 \end{array}\right) ,
\end{equation}
respectively, where these matrices are expressed in the computational basis $\{ |0\rangle ,|1\rangle \}$.  Let us now define three bases of $\mathbb{C}^{2}$, which will correspond to eigenstates of observables (see Figure 1).  We start with the basis $\{ |+x\rangle , |-x\rangle \}$, where $|\pm x\rangle = (|0\rangle \pm |1\rangle )/\sqrt{2}$.  In addition, define $\{ |u_{0}\rangle = U|+x\rangle ,\ |u_{1}\rangle =U|-x\rangle \}$ and
$\{ |v_{0}\rangle = U^{2}|+x\rangle ,\ |v_{1}\rangle =U^{2}|-x\rangle \}$.  Noting that $V|\pm x\rangle = |\mp x\rangle$, we see that if we apply the matrices corresponding to the remaining group elements, $V$, $UV$, or $U^{2}V$ to the states $|\pm x\rangle$ we will just obtain elements of one of the three bases we have just defined.

\begin{figure}[h!]
  \centering
   \includegraphics[width=0.3\textwidth]{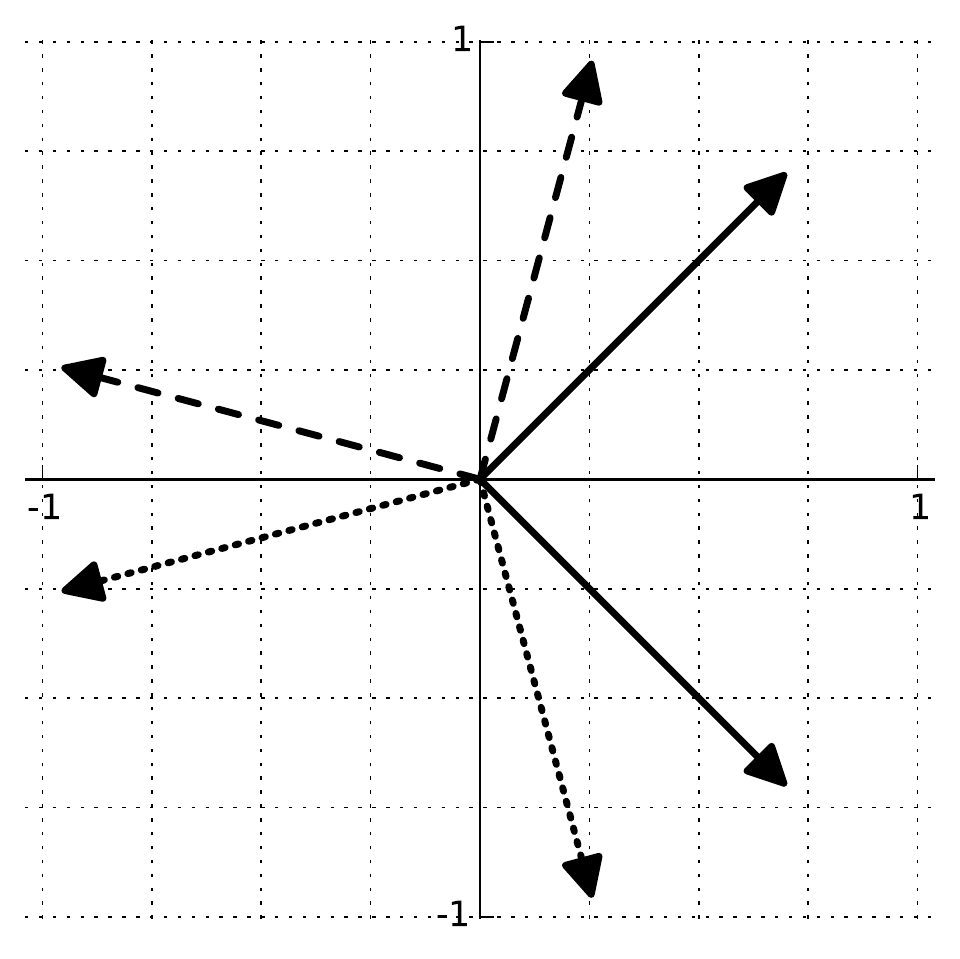} 
   \caption{The three different bases.  Solid is $|\pm x\rangle$, dashed is $\{ |u_{0}\rangle ,|u_{1}\rangle \}$, and dotted is $\{ |v_{0}\rangle , |v_{1}\rangle \}$ .}
   \label{di3_bases}
\end{figure}

We now consider a situation in which two parties, Alice and Bob, perform measurements on two qubits, each party possessing one of the qubits.  Each party can perform one of three measurements, and each measurement has two possible outcomes.  Alice's observables are $a_{0}$, $a_{1}$. and $a_{2}$, and Bob's are $b_{0}$, $b_{1}$, and $b_{2}$, and in each case the result of the measurement will be either $0$ or $1$.  In the case that the measurements are described by quantum mechanics, we will have $a_{0}=|-x\rangle\langle -x|$, $a_{1} =|u_{1}\rangle\langle u_{1}|$, and $a_{2}=|v_{1}\rangle\langle v_{1}|$, and similarly for the $b_{j}$, $j=0,1,2$. 

 Next we find a set of probabilities whose sum will give us a Bell inequality.  We begin with the representation of $D_{3}$ on $\mathbb{C}^{2}\otimes \mathbb{C}^{2}$ given by $g\in D_{3} \rightarrow \Gamma^{(3)}(g)\otimes \Gamma^{(3)}(g)$, where $\Gamma^{(3)}(g)$ is the matrix in the representation $\Gamma^{(3)}$ corresponding to the group element $g$ (this matrix will be a product of powers of the matrices $U$ and $V$).  Application of these matrices to elements of $\mathbb{C}^{2}\otimes \mathbb{C}^{2}$ gives us a group action, and we are going to be interested in particular orbits.  Starting with the state $|+x,+x\rangle$ and applying to it the matrices $\Gamma^{(3)}(g)\otimes \Gamma^{(3)}(g)$, for all $g\in D_{3}$, we have the orbit
\begin{equation}
\{ |+x,+x\rangle , |u_{0},u_{0}\rangle , |v_{0},v_{0}\rangle , |-x,-x\rangle , |u_{1},u_{1}\rangle , |v_{1},v_{1}\rangle \} .
\end{equation}
Projections onto these states correspond to the measurement probabilities for Alice and Bob
\begin{eqnarray}
\label{orbit-prob-1}
\{ p(a_{0}=0,b_{0}=0), p(a_{1}=0, b_{1}=0), \nonumber \\
p(a_{2}=0, b_{2}=0), p(a_{0}=1,b_{0}=1), \nonumber \\
p(a_{1}=1,b_{1}=1), p(a_{2}=1,b_{2}=1) \} .
\end{eqnarray}
We obtain a second orbit by starting with $|-x,v_{0}\rangle$,
\begin{equation}
\{ |-x,v_{0}\rangle , |u_{1},+x\rangle , |v_{1},u_{0}\rangle , |+x,u_{1}\rangle , |u_{0},v_{1}\rangle , |v_{0},-x\rangle \} .
\end{equation}
Projections onto these states correspond to the measurement probabilities 
\begin{eqnarray}
\label{orbit-prob-2}
\{ p(a_{0}=1,b_{2}=0), p(a_{1}=1, b_{0}=0), \nonumber \\
p(a_{2}=1, b_{1}=0), p(a_{0}=0,b_{1}=1), \nonumber \\
p(a_{1}=0,b_{2}=1), p(a_{2}=0,b_{0}=1) \} .
\end{eqnarray}

We now want to find the quantum state than maximizes the sum of the probabilities in Eqs.\ (\ref{orbit-prob-1}) and (\ref{orbit-prob-2}).  This can be accomplished by finding the largest eigenvalue of the operator
\begin{equation}
A=\sum_{g\in D_{3}} \left( \Gamma^{(3)}(g)\otimes \Gamma^{(3)}(g)\right) L \left( \Gamma^{(3)}(g)\otimes \Gamma^{(3)}(g)\right)^{\dagger} ,
\end{equation}
where $L=|+x,+x\rangle\langle +x,+x|+ |-x,v_{0}\rangle\langle -x,v_{0}|$, and its corresponding eigenstate.  The expectation value of $A$ in a state $|\phi\rangle \in \mathbb{C}^{2}\otimes \mathbb{C}^{2}$ is just the sum of the probabilities in Eqs.\ (\ref{orbit-prob-1}) and (\ref{orbit-prob-2}) for that state.

One can simply do a brute force calculation to find the largest eigenvalue of $A$, but it is also possible  to make use of group representation theory.  The representation $\Gamma^{(3)}\otimes \Gamma^{(3)}$ is reducible, and can be split into its irreducible components
\begin{equation}
\Gamma^{(3)}\otimes \Gamma^{(3)} = \Gamma^{(1)}\oplus \Gamma^{(2)}\oplus \Gamma^{(3)} ,
\end{equation}
where each irreducible component acts on an invariant subspace.  This follows from the relation \cite{cornwell}
\begin{equation}
\label{decomposition}
n_{p}=\frac{1}{|G|} \sum_{g\in G}\chi(g) \chi^{(p)}(g)^{\ast} ,
\end{equation}
which gives the number of times, $n_{p}$, an irreducible representation of a group $G$, $\Gamma^{(p)}$ appears in the decomposition of a representation $\Gamma$.  Here $|G|$ is the order (number of elements) of $G$, $\chi (g)$ is the character of $\Gamma (g)$, and $\chi^{(p)}(g)$ is the character of $\Gamma^{(p)}(g)$ 
We find that the space corresponding to $\Gamma^{(1)}$ is spanned by $(|00\rangle + |11\rangle )/\sqrt{2}$, the space corresponding to $\Gamma^{(2)}$ is spanned by $(|01\rangle - |10\rangle )/\sqrt{2}$, and the space corresponding to $\Gamma^{(3)}$ is spanned by $(|00\rangle - |11\rangle )/\sqrt{2}$ and $(|01\rangle + |10\rangle )/\sqrt{2}$.  

We can now make use of the following relation \cite{cornwell}.  If $G$ is a group and $\Gamma^{(p)}$ and $\Gamma^{(q)}$ are irreducible representations of $G$, then
\begin{equation}
\label{sum-two-irreps}
\frac{1}{|G|} \sum_{g\in G} \Gamma^{(p)}(g)_{jk}^{\ast} \Gamma^{(q)}(g)_{st} = \frac{1}{d_{p}}\delta_{pq} \delta_{js} \delta_{kt} ,
\end{equation}
where $d_{p}$ is the dimension of the irreducible representation $\Gamma^{(p)}$.  Now let $\{ |\alpha_{j}^{(p)}\rangle \}$ be an orthonormal basis of a carrier space for the irreducible representation $\Gamma^{(p)}$, $|X_{p}\rangle$ a vector in that space, $\{ |\beta_{j}^{(q)}\rangle \}$ an orthonormal basis for a carrier space for the irreducible representation $\Gamma^{(q)}$, and $|X_{q}\rangle$ a vector in that space.  We then have that
\begin{eqnarray}
\frac{1}{|G|} \sum_{g\in G}\langle \alpha_{j}^{(p)}|\Gamma^{(p)}(g)|X_{p}\rangle \langle X_{q}|\Gamma^{(q)\dagger}(g)|\beta_{j^{\prime}}^{(q)}\rangle \nonumber \\
= \frac{1}{|G|} \sum_{g\in G} \sum_{k,k^{\prime}} \langle \alpha_{j}^{(p)}|\Gamma^{(p)}(g)|\alpha_{k}^{(p)}\rangle  \langle \beta_{j^{\prime}}^{(q)}| \Gamma^{(q)}(g)|\beta_{k^{\prime}}^{(q)}\rangle^{\ast} \nonumber \\
\langle \alpha_{k}^{(p)}|X_{p}\rangle \langle X_{q}|\beta_{k^{\prime}}^{(q)}\rangle .
\end{eqnarray}
Making use of Eq.\ (\ref{sum-two-irreps}) this becomes
\begin{equation}
\label{two-irreps-final}
\frac{1}{d_{p}}\delta_{pq}\delta_{jj^{\prime}}\sum_{k} \langle X_{q}|\alpha_{k}^{(q)}\rangle  \langle\beta_{k}^{(p)}|X_{p}\rangle .
\end{equation}
If, in the case $p=q$ the carrier spaces are the same, this reduces to
\begin{equation}
\label{same-carrier}
\frac{1}{d_{p}}\delta_{pq}\delta_{jj^{\prime}} \| X_{p}\|^{2} .
\end{equation}
Note that if when $p=q$ and the carrier spaces are not the same, we are assuming that the basis elements $|\alpha_{j}^{(p)}\rangle$ and $|\beta_{j}^{(p)}\rangle$ transform in the same way under the action of $\Gamma^{(p)}$.  Finally, suppose we have a representation of $G$, $\Gamma (g)$, which is the direct sum of irreducible representations each of which only appears once.  In this case we will have that $|\alpha_{j}^{(p)}\rangle = |\beta_{j}^{(p)}\rangle$ when $p=q$, so that we can make use of Eq.\ (\ref{same-carrier}).  If we then have a vector
\begin{equation}
|\psi\rangle = \sum_{q,j} c_{q,j}|\alpha_{j}^{(q)}\rangle = \sum_{q} |\psi_{q}\rangle ,
\end{equation}
where $|\psi_{q}\rangle = \sum_{j} c_{q,j} |\alpha_{j}^{(q)}\rangle$ is the component of $|\psi\rangle$ that is in the subspace that transforms according to $\Gamma^{(q)}$, then
\begin{equation}
\frac{1}{|G|} \sum_{g\in G} \Gamma (g)|X\rangle\langle X|\Gamma^{\dagger}(g)|\psi\rangle = \sum_{p} \frac{1}{d_{p}} \| X_{p}\|^{2} |\psi_{p}\rangle ,
\end{equation}
where $X_{p}$ is the projection of $|X\rangle$ onto the subspace that transforms according to $\Gamma^{(p)}$.

Now let us apply this to find the eigenstates of $A$.  Setting $|X^{(1)}\rangle = |+x,+x\rangle$ and $|X^{(2)}\rangle = |-x,v_{0}\rangle$, we have that 
\begin{eqnarray}
A|\psi\rangle & = & 6\sum_{p=1}^{2} (\| X_{p}^{(1)}\|^{2} + \| X_{p}^{(2)}\|^{2}) |\psi_{p}\rangle \nonumber \\
& & + 3(\| X_{3}^{(1)}\|^{2} + \| X_{3}^{(2)}\|^{2}) |\psi_{3}\rangle .
\end{eqnarray}
From this we see that the eigenvectors of $A$ are just vectors lying in the invariant subspaces, and the eigenvalues are
\begin{eqnarray}
6(\| X_{1}^{(1)}\|^{2} + \| X_{1}^{(2)}\|^{2}) & = & \frac{21}{4} \nonumber \\
6(\| X_{2}^{(1)}\|^{2} + \| X_{2}^{(2)}\|^{2}) & = & \frac{3}{4} \nonumber \\
3(\| X_{3}^{(1)}\|^{2} + \| X_{3}^{(2)}\|^{2}) & = & 3 .
\end{eqnarray}
Therefore, the largest eigenvalue is $21/4$ and the corresponding eigenvector is $(|00\rangle + |11\rangle )/\sqrt{2}$.

The classical bound on the sum of the $12$ probabilities is found as before.  We assume that the probabilities can be derived from a joint distribution, $P(a_{0},b_{0}; a_{1},b_{1};a_{2},b_{2})$ and calculate their sum in terms of the joint distribution.  The largest coefficient multiplying a probability from the joint distribution gives the upper bound to the sum, and in this case it is $5$.  Because $21/4 > 5$, the quantum result violates the classical inequality. 

\subsection{Comparison of nonlocal games}
The Bell inequality derived in the last section can be rephrased in terms of a nonlocal game, and we shall do so shortly.  It is useful, however to compare that nonlocal game to one resulting from a Bell inequality that is produced by an abelian group, in particular the group $\mathbb{Z}_{6}$. 
 
The group $\mathbb{Z}_{6}$ has a single generator whose sixth power is just the identity element.  We will choose the representation of $\mathbb{Z}_{6}$ generated by the matrix, $U$, on $\mathbb{C}^{2}$ (qubits) given by
 \begin{equation}
 U=|+x\rangle\langle +x|+e^{i\pi /3}|-x\rangle\langle -x| .
 \end{equation}
 Note that $U^{6}=I$ and $U^{3}=|+x\rangle\langle +x| - |-x\rangle\langle -x| =\sigma_{x}$.  We can use $U$ to define three bases, the computational basis $\{ |0\rangle , |1\rangle \}$, $\{ |u_{j}\rangle =U|j\rangle | j=0,1\}$, and $\{ |v_{j}\rangle = U^{2}|j\rangle | j=0,1 \}$.  These bases are the eigenstates of three observables, $a_{0}=|1\rangle\langle 1|$, $a_{1}=|u_{1}\rangle\langle u_{1}|$, and $a_{2}=|v_{1}\rangle\langle v_{1}|$.  
 
 We now consider the tensor product space $\mathbb{C}^{2}\otimes \mathbb{C}^{2}$ (two qubits) and the representation of $\mathbb{Z}_{6}$ generated by $U\otimes U$.  The observables on the first qubit are $a_{j}$ for $j=0,1,2$, and we denote the identical observables on the second qubit by $b_{j}$, for $j=0,1,2$.  We now look at two orbits.  The first starts with the state $|0,0\rangle$ and is given by
 \begin{equation}
 \{ |0,0\rangle , |u_{0}, u_{0}\rangle , |v_{0}, v_{0}\rangle , |1,1\rangle , |u_{1}, u_{1}\rangle , |v_{1}, v_{1}\rangle \} ,
 \end{equation}
 corresponding to the measurement probabilities
\begin{eqnarray}
\label{orbit-prob-1-z6}
\{ p(a_{0}=0,b_{0}=0), p(a_{1}=0, b_{1}=0), \nonumber \\
p(a_{2}=0, b_{2}=0), p(a_{0}=1,b_{0}=1), \nonumber \\
p(a_{1}=1,b_{1}=1), p(a_{2}=1,b_{2}=1) \} .
\end{eqnarray} 
The second orbit begins with the state $|0,u_{0}\rangle$ and is 
\begin{equation}
 \{ |0,u_{0}\rangle , |u_{0}, v_{0}\rangle , |v_{0}, 1\rangle , |1,u_{1}\rangle , |u_{1}, v_{1}\rangle , |v_{1}, 0\rangle \} ,
 \end{equation}
which corresponds to the measurement probabilities
\begin{eqnarray}
\label{orbit-prob-2-z6}
\{ p(a_{0}=0,b_{1}=0), p(a_{1}=0, b_{2}=0), \nonumber \\
p(a_{2}=0, b_{0}=1), p(a_{0}=1,b_{1}=1), \nonumber \\ 
p(a_{1}=1,b_{2}=1), p(a_{2}=1,b_{0}=0) \} .
\end{eqnarray} 

We next want to find the quantum state that maximizes the sum of the probabilities in Eqs.\ (\ref{orbit-prob-1-z6}) and (\ref{orbit-prob-2-z6}).  As before, it will be the eigenstate of the operator
\begin{equation}
A=\sum_{j=0}^{5} (U\otimes U)^{j}(|0,0\rangle\langle 0,0| + |0,u_{0}\rangle\langle 0,u_{0}|) (U^{\dagger}\otimes U^{\dagger})^{j} ,
\end{equation}
corresponding to the largest eigenvalue.  The eigenstates of $U\otimes U$ will also be the eigenstates of $A$ \cite{guney}, and the eigenvalues of $U\otimes U$ are $1$ and $e^{2\pi i/3}$, which are non-degenerate, and $e^{i\pi /3}$, which is doubly degenerate.  We find that the eigenstate of $A$ with the largest eigenvalue lies in the space corresponding to the $e^{i\pi /3}$ eigenvalue of $U\otimes U$, and this space is spanned by the vectors $|+x, -x\rangle$ and $|-x, +x\rangle$.  In this space, $A$ reduces to the $2\times 2$ matrix
\begin{equation}
\frac{1}{4}\left( \begin{array}{cc} 1 & 1+e^{i\pi /3} \\ 1+e^{-i\pi /3} & 2 \end{array}\right) ,
\end{equation}
and the largest eigenvalue is $3 + (3/2)\sqrt{3}$ with the corresponding eigenvector
\begin{equation}
\label{phi-z6}
|\phi\rangle = \frac{1}{\sqrt{6}}[ (1+e^{i\pi /3})|+x,-x\rangle + \sqrt{3}|-x,+x\rangle ] .
\end{equation}
The maximum classical value of the sum of the probabilities in Eqs.\ (\ref{orbit-prob-1-z6}) and (\ref{orbit-prob-2-z6}), that is the sum if all of the probabilities come from a joint distribution of all six observables is $5$.  Since $3 + (3/2)\sqrt{3} > 5$, the quantum probabilities violate the classical bound.
 
We can now proceed to a discussion of this Bell inequality in terms of a nonlocal game.  The two players are Alice and Bob, and an arbitrator sends Alice a value of $s\in \{0,1,2\}$ and Bob a value of $t\in \{0,1,2\}$.  Not all values of $(s,t)$ are allowed.  In particular, either $s=t$ or $(s,t)$ must be $(0,1)$, $(1,2)$, or $(2,0)$, so that six out of the nine possibilities are allowed, and they will be assumed to be equally probable.  Alice and Bob then each send a bit to the arbitrator.  They win if their bit values differ and $(s,t)=(2,0)$ or their bit values are the same and $(s,t)$ is any of the other allowed values.  Note that for each allowed value of $(s,t)$ there are two winning possibilities.  Classically their winning probability is $5/6$, and it can be achieved if Alice and Bob each always send the bit value $0$.  In the quantum case, Alice and Bob share the state $|\phi\rangle$ in Eq.\ (\ref{phi-z6}), and the values of $(s,t)$ determine which observable they measure, in particular, Alice measures $a_{s}$ and Bob measures $b_{t}$.  The bit values they send to the arbitrator are simply the results of their measurements.  In this scenario, all of the probabilities in Eqs.\ (\ref{orbit-prob-1-z6}) and (\ref{orbit-prob-2-z6}) are the same, and are equal to $(2+\sqrt{3})/8$.  For each value of $(s,t)$ there are two winning possibilities, so the overall winning probability for Alice and Bob using the quantum strategy is $2(2+\sqrt{3})/8 = (2+\sqrt{3})/4$.  Comparing the classical and quantum strategies, we see that the best classical strategy has a winning probability of approximately $0.83$ while the quantum strategy has a winning probability of $0.93$, so there is a quantum advantage.

Now let us go back and rephrase the Bell inequality that resulted from $D_{3}$ as a nonlocal game.  As we shall see, its structure is different than that of the game that resulted from $\mathbb{Z}_{6}$.  The basic situation is as before, Alice receives a value of $s\in \{0,1,2\}$ and Bob receives a value of $t\in \{0,1,2\}$, but in this case all nine combinations of $(s,t)$ are possible.   Alice and Bob then send a bit to an arbitrator.  They win if $s=t$ and they return the same bit value or if $s$ and $t$ are different, they return the bit values $(a,b)$ that are shown in Table 3.  Note that in this case when $s=t$ there are two winning possibilities for $(a,b)$, but for $s\neq t$ there is only one.  This is different from the previous game where for each allowed value of $(s,t)$ there were two winning possibilities.  

\begin{table}
\label{win-values}
\centering
\begin{tabular}{|c|c|}\hline
(s,t) & (a,b) \\ \hline (0,1) & (0,1) \\ (1,0) & (1,0) \\ (0,2) & (1,0) \\ (2,0) & (0,1) \\ (1,2) & (0,1) \\ (2,1) & (1,0)
\\ \hline
\end{tabular} 
\caption{Winning values for the nonlocal game derived from $D_{3}$ when $s\neq t$. }
\end{table}

Now let us look at the classical and quantum winning probabilities.  .  As we saw, the sum of the probabilities resulting from $D_{3}$ is $5$, and by an argument similar to that in Section 2 that implies that the classical winning probability is less than or equal to $5/9\simeq 0.556$.  This bound can be achieved with the following strategy.  If Alice receives $s$ from the arbitrator, she returns $f_{A}(s)$, and when Bob receives $t$, he returns $f_{B}(t)$, where $f_{A}(0)=f_{B}(0)=1$, $f_{A}(1)=f_{B}(1)=0$, and $f_{A}(2)=f_{B}(2)=0$.  In the quantum case, Alice and Bob share the state $|\phi\rangle = |00\rangle + |11\rangle )/\sqrt{2}$, and Alice measures $a_{s}$ and Bob measures $b_{t}$, where $a_{s}$ and $b_{t}$ are the observables appropriate for the Bell inequality that resulted from $D_{3}$, i.e.\ $a_{0}=|-x\rangle\langle -x|$, $a_{1} =|u_{1}\rangle\langle u_{1}|$, and $a_{2}=|v_{1}\rangle\langle v_{1}|$, and similarly for the $b_{j}$, $j=0,1,2$.  They then report their measurement results as their bit values.  The quantum winning probability is then just $1/9$ times the sum of the probabilities in Eqs.\ (\ref{orbit-prob-1}) and (\ref{orbit-prob-2}), which is $7/12 \simeq 0.583$.  This is larger than the classical winning probability.  In this case, it is worth noting that the quantum probabilities for the two different orbits are not the same.  For the case that $s=t$, i.e. the probabilities in Eq.\ (\ref{orbit-prob-1}), the probabilities are all equal to $1/2$.  This implies that if $s=t$, Alice and Bob always win the game if they are using the quantum strategy.  When $s\neq t$, the probabilities in Eq.\ (\ref{orbit-prob-2}), are all equal to $3/8$.

To summarize the results of this section, we have constructed two nonlocal games, one based on $\mathbb{Z}_{6}$ and the other based on $D_{3}$.  For both, Alice received a value of $s$ and Bob received a value of $t$, and each had to return a bit value.  In one game, the set of allowed values of $(s,t)$ was restricted, in the other it was not.  In addition, in one game for each allowed value of $(s,t)$ there were always two winning values of $(a,b)$, while in the second game this was true if $s=t$, but there was only one winning value otherwise.  Therefore, the nonlocal games had rather different structures.

 \subsection{$D_{6}$}
To conclude we will look at a larger group, $D_{6}$.  This group has the generators $r$ and $s$, where $s^{2}=e$, as before, but now $r^{6}=e$.  This group has six conjugacy classes: $C_{e}=\{ e\}$, $C_{r}=\{ r,r^{5}\}$, $C_{r^{2}}=\{ r^{2}, r^{4}\}$, $C_{r^{3}}=\{ r^{3}\}$, $C_{s}=\{ s, r^{2}s, r^{4}s \}$, and $C_{rs}=\{ rs, r^{3}s, r^{5}s \}$.   It has six irreducible representations, $\Gamma^{(j)}$ for $j=1,2,3,4$, which are one-dimensional, and $\Gamma^{(5)}$ and $\Gamma^{(6)}$, which are two-dimensional.  The character table for this group is given in Table 4.
 
\begin{table}
\centering
\begin{tabular}{|c|c|c|c|c|c|c|} \hline
 & $C_{e}$ & $C_{r}$ & $C_{r^{2}}$ & $C_{r^{3}}$ & $C_{s}$ & $C_{rs}$ \\ \hline $\Gamma^{(1)}$ & $1$ & $1$& $1$& $1$ & $1$& $1$ \\ \hline $\Gamma^{(2)}$ & $1$ & $1$ & $1$ & $1$ & $-1$ & $-1$ \\ \hline $\Gamma^{(3)}$ & $1$ & $-1$ & $1$ & $-1$ & $1$ & $-1$ \\ \hline $\Gamma^{(4)}$ & $1$ & $-1$ & $1$ & $-1$ & $-1$ & $1$ \\ \hline $\Gamma^{(5)}$ & $2$ & $1$ & $-1$ & $-2$ & $0$ & $0$ \\ \hline $\Gamma^{(6)}$ & $2$ & $-1$ & $-1$ & $2$ & $0$ & $0$ \\ \hline
\end{tabular}
\caption{Character table for $D_{6}$.}
\end{table}
 
We will make use of the following representation of $D_{6}$ on $\mathbb{C}^{3}$.  The computational basis is $\{ |j\rangle | j=0,1,2\}$, and let us define another basis 
\begin{equation}
|u_{j}\rangle = \frac{1}{\sqrt{3}}\sum_{k=0}^{2} e^{2\pi ijk/3} |k\rangle .
\end{equation}
Corresponding to the group element $r$, we choose
\begin{equation}
U=|u_{0}\rangle\langle u_{0}| + e^{-i\pi /3}|u_{1}\rangle\langle u_{1}| + e^{i\pi /3}|u_{2}\rangle\langle u_{2}| ,
\end{equation}
and corresponding to $s$ we choose
\begin{equation}
V=|u_{0}\rangle\langle u_{0}| + i(|u_{1}\rangle\langle u_{2}| - |u_{2}\rangle\langle u_{1}|) .
\end{equation}
This choice for $U$ was used in a previous paper as a generator of a representation of $\mathbb{Z}_{6}$ \cite{guney}.  Note that it has the property that $U^{2}|j\rangle = |j+1\rangle$, where the addition is modulo $3$.  If we denote the representation generated by $U$ and $V$ by $\Gamma$, then application of Eq.\ (\ref{decomposition}) gives us that
\begin{equation}
\Gamma = \Gamma^{(1)} \oplus \Gamma^{(5)} .
\end{equation}

Application of powers and products of the operators $U$ and $V$ to the computational basis yield three additional bases, $\{ |v_{j}\rangle = U|j\rangle | j=0,1,2\} $, $\{ |w_{j}\rangle = V|j\rangle | j=0,1,2\}$, and $\{ |x_{j}\rangle = UV|j\rangle | j=0,1,2\}$.  We can now define four observables that take values in the set $\{ 0,1,2\}$
\begin{eqnarray}
a_{0}= \sum_{j=1}^{2} j|j\rangle\langle j| & &a_{1}=\sum_{j=1}^{2}j|v_{j}\rangle\langle v_{j}| \nonumber \\
a_{2}=\sum_{j=1}^{2}j|w_{j}\rangle\langle w_{j}| & &a_{3}=\sum_{j=1}^{2}j|x_{j}\rangle\langle x_{j}| . 
\end{eqnarray}
In the bipartite case, Alice and Bob will choose among these observables for their measurements.  That is, they each possess a qutrit and decide to measure one of the four observables above (we will denote Bob's observables by $b_{j}$).

Next we will choose two orbits.  These were again identified by means of a random search that checked for Bell inequality violations.  The orbits start on the states $(U^{4}\otimes U^{2}V|0,0\rangle = |2, w_{2}\rangle$ and $(I\otimes U^{5}V)|0,0\rangle = |0,x_{1}\rangle$ and further elements of the orbits are found by applying $\Gamma (g)\otimes \Gamma (g)$, for $g\in D_{6}$, to the initial states.  Each orbit contains $12$ states and gives rise to $12$ corresponding measurement probabilities.  The $24$ probabilities that result from these two orbits are listed in the Appendix B.  If all of these probabilities come from a joint distribution, their sum cannot be greater than $6$.

We now need to see if we can find a quantum state that violates the classical bound.  The operator $A$ is now given by
\begin{equation}
\label{d6-A}
A=\sum_{g\in D_{6}} \left( \Gamma (g)\otimes \Gamma (g)\right) L \left( \Gamma (g)\otimes \Gamma (g)\right)^{\dagger} ,
\end{equation}
where now $L=|2,w_{2}\rangle\langle 2,w_{2}|+ |0,x_{1}\rangle\langle 0,x_{1}|$.  Application of Eq.\ (\ref{decomposition}) gives us that
\begin{equation}
\Gamma \otimes \Gamma =2\Gamma^{(1)} \oplus \Gamma^{(2)} \oplus 2\Gamma^{(5)}\oplus \Gamma^{(6)} .
\end{equation}
Using the representation of $U$ and $V$ in the $\{ |u_{j}\rangle \}$ basis, we find that $|u_{0}, u_{0}\rangle$ and $(|u_{1}, u_{2}\rangle + |u_{2}, u_{1}\rangle )/\sqrt{2}$ transform as $\Gamma^{(1)}$,  $(|u_{1}, u_{2}\rangle - |u_{2}, u_{1}\rangle )/\sqrt{2}$ transforms as $\Gamma^{(2)}$, both $\{ |u_{0}, u_{1}\rangle ,  |u_{0}, u_{2}\rangle \}$ and $\{ |u_{1}, u_{0}\rangle , |u_{2}, u_{0} \rangle \}$ transform as $\Gamma^{(5)}$, and $\{ |u_{1}, u_{1}\rangle , |u_{2} , u_{2}\rangle \}$ transform as $\Gamma^{(6)}$. 

We find that the largest eigenvalue of $A$ corresponds to an eigenvector that lies in the subspace spanned by the two vectors that transform as $\Gamma^{(1)}$.  The details of the calculation are in  Appendix B, but here we note the following.  According to Eq.\ (\ref{sum-two-irreps}), the eigenvectors of $A$ will lie in invariant subspaces corresponding to the representations $\Gamma^{(1)}$, $\Gamma^{(2)}$, $\Gamma^{(5)}$, and $\Gamma^{(6)}$, and these subspaces are orthogonal.  The components of both $|2,w_{2}\rangle$ and $|0,x_{1}\rangle$ that lie in the $\Gamma^{(1)}$ subspace are the same, and are given by 
\begin{equation}
|X_{1}\rangle = \frac{1}{3}|u_{0},u_{0}\rangle - \frac{1}{2\sqrt{3}}(|u_{1},u_{2}\rangle + |u_{2},u_{1}\rangle ) .
\end{equation}  
Because it transforms as $\Gamma^{(1)}$, this vector is invariant under the actions of $U$ and $V$, and this implies that in the $\Gamma^{(1)}$ space, $A$ is just $2(12)|X_{1}\rangle\langle X_{1}|$.  Therefore, the two eigenvectors of $A$ in this subspace are the vector orthogonal to $|X_{1}\rangle$, which has an eigenvalue of $0$, and a normalized version of $|X_{1}\rangle$, which is $|\phi\rangle = 3(\sqrt{2/5})|X_{1}\rangle$, whose eigenvalue is $2(12)\| X_{1}\|^{2}=20/3$.  As $(20/3) > 6$, the sum of the probabilities for the state $|\phi\rangle$ violates the classical bound, so the sum of the $24$ probabilities in Table 5 gives us a Bell inequality.

\section{Conclusion}
We have shown how group actions can be used to generate Bell inequalities.  In particular, we provided an example of a three-party Bell inequality using an Abelian group, and two examples of two-party inequalities but with non-Abelian groups.  The orbits of the group action are used to generate events, the sum of whose probabilities is the main object appearing in the Bell inequality.  This approach has the benefit of providing a set of quantum observables that can be measured to test the Bell inequality and a quantum state that violates it.

There are a number of areas in which the research presented here could be extended.  The choice of the orbits that led to the Bell inequalities was done by using a random search (see Appendix A).  It would be useful to have a criterion for choosing them.  This would also allow us to gain a better understanding of how the structures of Bell inequalities are related to the underlying groups.  The Bell inequalities depend on both the group and the choice of orbits, and at the moment we do not have a good way of disentangling these two effects.  A better understanding of how to choose the orbits would, we hope, lead to a better idea of the relation between the group and the Bell inequality.  

\section*{Acknowledgment} 
This research was supported by a grant from the John Templeton Foundation. 

\section*{Appendix A}
Here we provide more detail about how the random search to determine the orbits for the group $D_{3}$ was performed.  The two orbits that yield a Bell inequality were found by a random search in the space of all possible orbit pairs made with the SAGE (\url{http://www.sagemath.org/}), an open source computer algebra system. SAGE includes group theory and symbolic manipulation packages that are suitable for this task.

First, using SAGE the $D_3$ group is generated and its elements $g\in{}D_3$ are calculated. Then the group generators $r$ and $s$ are associated with the corresponding representation matrices, $U=\Gamma{}\left(r\right)$, $V=\Gamma{} \left( s \right)$. We know how the rest of the group elements are generated from the generators $\{g_i | i=0,1,\ldots 5\}=\{ e,r,r^{2}, s, rs, r^{2}s \}$. The associated representation matrices are calculated accordingly, $\{ \Gamma\left( g_i \right) = \Gamma_i | i=0,\ldots 5 \} = \{ I, U, U^2, V, UV, U^2V \}$. To associate the representation matrices with quantum measurement outcome states, the matrices are applied to a chosen initial state, which in our case was $|+x\rangle$, giving $|\psi_i\rangle = \Gamma_i |+x\rangle$. 

In the code, then, the orthogonality relations among these states are analyzed. A table $T_{ij} = |\langle \psi_i | \psi_j \rangle|$ of the absolute values of inner products is calculated. From the table these states are classified into different orthonormal bases, with each basis corresponding to the  different possible eigenstates of a single observable. States for which the inner products are $0$ or $1$ are in the same basis. In this way the Bell scenario for the number of measurements and outcomes is determined. The choice of initial state is essential to be able to get useful orthonormal bases.  To be specific, each state is associated with an event $E$, namely an observable and its outcome,  $|\psi_i\rangle \leftrightarrow a_{m(g_i)}=o(g_i)$, where $m$ is the choice of observable, and $o$ is the outcome. For our choice of $U$, $V$ and the initial state, $|+x\rangle$, the 6 states $|\psi_i \rangle$ belong to 3 two dimensional orthonormal bases. $\{ E_i | i=0, \ldots 5 \} = \{ a_{0}=0, a_{1}=0, a_{2}=0, a_{0}=1, a_{1}=1, a_{2}=1\}$.

We have two parties, and we want to see whether two orbits are sufficient. For each orbit we need two group elements, $g_\mu$ and $g_\nu$, to set the initial joint state $|\Psi_{\mu,\nu}\rangle = \Gamma(g_\mu) |+x\rangle \otimes \Gamma(g_\nu) |+x\rangle$. Then, the orbit will give us the $A$ operator 
\begin{equation}
A_{\mu,\nu} = \sum_i \left( \Gamma(g_i)\otimes \Gamma(g_i)\right) |\Psi_{\mu,\nu}\rangle \langle \Psi_{\mu,\nu}| \left( \Gamma(g_i)^{\dagger}\otimes \Gamma^\dagger{}(g_i) \right).
\end{equation}
The $A$ corresponding to both orbits is $A = A_{\mu_1,\nu_1} + A_{\mu_2,\nu_2}$.  The choice of $\mu_1,\nu_1, \mu_2,\nu_2$ also determines the set of joint probabilities
\begin{eqnarray}
\mathcal{P} & = & \left\{ P\left( a_{m(g_i  g_{\mu_j})} =o(g_i  g_{\mu_j}), b_{n(g_i  g_{\nu_j})} = o(g_i  g_{\nu_j})  \right) \right. \nonumber \\
& & \left. |i=0,\ldots 5, j=1,2 \right\}.
\end{eqnarray}

Because the size of the search space increases exponentialy with respect to the group size a random search is implemented. The size is $|G|^{N_o N_p}$ where $|G|$ is the order of the group, $N_o$ is the number of orbits we want, and $N_p$ is the number of parties. For a random choice of $\{ \mu_1,\nu_1, \mu_2,\nu_2 \}$ the biggest eigenvalue of $A$, $\lambda_{max}$, is compared with the classical bound of the sum of the joint probabilities in $\mathcal{P}$, $c$. A violation is found when $\lambda_{max} > c$. The code can be downloaded from \url{http://www.github.com/vug/bell-group-actions} .

\section*{Appendix B}
Table 5 gives the probabilities corresponding to the two orbits for $D_{6}$.  The starting states for the orbits are given in the first line, and the group representation element that is applied to the initial state to give the resulting probability is given in the first column.
\begin{table}
\centering
\begin{tabular}{|c|c|c|}\hline & $|2,w_{2}\rangle$ & $|0,x_{1}\rangle$ \\ \hline $I$ & $p(a_{0}=2, b_{2}=2)$ & $p(a_{0}=0, b_{3}=1)$ \\ $U$ & $p(a_{1}=2, b_{3}=2)$ & $p(a_{1}=0, b_{2}=0)$ \\ $U^{2}$ & $p(a_{0}=0, b_{2}=1)$ & $p(a_{0}=1, b_{3}=0)$ \\ $U^{3}$ & $p(a_{1}=0, b_{3}=1)$ & $p(a_{1}=1, b_{2}=2)$ \\ $U^{4}$ & $p(a_{0}=1, b_{2}=0)$ & $p(a_{0}=2, b_{3}=2)$ \\ $U^{5}$ & $p(a_{1}=1, b_{3}=0)$ & $p(a_{1}=2, b_{2}=1)$ \\ $V$ & $p(a_{2}=2, b_{0}=2)$ & $p(a_{2}=0, b_{1}=0)$ \\ $UV$ & $p(a_{3}=2, b_{1}=2)$ & $p(a_{3}=0, b_{0}=1)$ \\ $U^{2}V$ & $p(a_{2}=1, b_{0}=0)$ & $p(a_{2}=2, b_{1}=1)$ \\ $U^{3}V$ & $p(a_{3}=1, b_{1}=0)$ & $p(a_{3}=2, b_{0}=2)$ \\ $U^{4}V$ & $p(a_{2}=0, b_{0}=1)$ & $p(a_{2}=1, b_{1}=2)$ \\ $U^{5}V$ & $p(a_{3}=0, b_{1}=0)$ & $p(a_{3}=1, b_{0}=0)$ \\ \hline
\end{tabular}
\caption{Probabilities generated by orbits for $D_{6}$.}
\end{table}

Next, we move on to the calculation of the eigenvalues and eigenstates of the operator $A$ for the group $D_{6}$ given in Eq.\ (\ref{d6-A}) .  The eigenvalues and eigenvectors for the $\Gamma^{(1)}$ subspace have already been discussed in the text.  The $\Gamma^{(2)}$ and $\Gamma^{(6)}$ subspaces are straightforward, since these representations only appear once in the decomposition of $\Gamma$ and we can use Eq.\ (\ref{same-carrier}).  For $\Gamma^{(2)}$ we find that the component of $|2,w_{2}\rangle$ in this subspace is 
\begin{equation}
|X_{2}^{(1)}\rangle = \frac{-i}{6}(|u_{1},u_{2}\rangle - |u_{2},u_{1}\rangle )
\end{equation}
while the component of $|0,x_{1}\rangle$ is just $|X_{2}^{(2)}\rangle = - |X_{2}^{(1)}\rangle$.  The eigenvalue corresponding to the $\Gamma^{(2)}$ space is then $4/3$.  The components of $|2,w_{2}\rangle$ and $|0,x_{1}\rangle$ in the $\Gamma^{(6)}$ subspace are
\begin{eqnarray}
|X_{6}^{(1)}\rangle & = & \frac{i}{3} (-|u_{1},u_{1}\rangle + |u_{2},u_{2}\rangle ) \nonumber \\
|X_{6}^{(2)}\rangle & = & -\frac{1}{3\sqrt{3}}[(1-e^{-2\pi i /3}) |u_{1},u_{1}\rangle \nonumber \\
& & + (1-e^{2\pi i/3}) |u_{2},u_{2}\rangle ] ,
\end{eqnarray}
respectively.  This gives $8/3$ as the eigenvalue corresponding to $\Gamma^{(6)}$, and this eigenvalue is two-fold degenerate.

The $\Gamma^{(5)}$ subspace is more complicated.  It is four dimensional and consists of two copies of the $\Gamma^{(5)}$ irreducible representation.  We first note that because $|u_{0}\rangle$ is invariant under the actions of $U$ and $V$, the states $|u_{0},u_{1}\rangle$ and $|u_{1},u_{0}\rangle$ transform in the same way, and the states $|u_{0},u_{2}\rangle$ and $|u_{2},u_{0}\rangle$ transform in the same way.  Now suppose that $|X_{5}\rangle$ is a vector in the $\Gamma^{(5)}$ subspace.  Setting $|\alpha_{j}\rangle = |u_{0},u_{j}\rangle$ and $|\beta_{j}\rangle = |u_{j},u_{0}\rangle$, for $j=1,2$, we find from Eq.\ (\ref{two-irreps-final}), that
\begin{eqnarray}
\sum_{g\in D_{6}} \Gamma (g)|X_{5}\rangle\langle X_{5}|\Gamma^{\dagger}(g) \nonumber \\
=6\left( \begin{array}{cccc} \|X_{5\alpha}\|^{2} & 0 & z & 0 \\ 0 & \|X_{5\alpha}\|^{2} & 0 & z \\ z^{\ast} & 0 & \|X_{5\beta}\|^{2} & 0 \\ 0 & z^{\ast} & 0 & \|X_{5\beta}\|^{2} \end{array} \right) ,
\end{eqnarray}
where the matrix is in the $\{ \alpha_{1}, \alpha_{2}, \beta_{1}, \beta_{2} \}$ basis, and 
\begin{eqnarray}
\|X_{5\alpha}\|^{2} & = & \sum_{j=1}^{2} |\langle X_{5}|\alpha_{j}\rangle |^{2} \nonumber \\
\|X_{5\beta}\|^{2}   & = & \sum_{j=1}^{2} |\langle X_{5}|\beta_{j}\rangle |^{2} \nonumber \\
z & = & \sum_{j=1}^{2} \langle X_{5}|\beta_{j}\rangle \langle \alpha_{j}|X_{5}\rangle .
\end{eqnarray}
The component of $|2,w_{2}\rangle$ transforming as $\Gamma^{(5)}$ is
\begin{eqnarray}
|X_{5}^{(1)}\rangle & = & \frac{1}{3\sqrt{3}}[ (1-e^{-2\pi i/3})|u_{0},u_{1}\rangle \nonumber \\
& & + (1-e^{2\pi i/3})|u_{0},u_{2}\rangle \nonumber \\
& & + \frac{1}{3}(e^{-2\pi i /3}|u_{1},u_{0}\rangle + e^{2\pi i/3}|u_{2},u_{0}\rangle ) ,
\end{eqnarray}
and the component of $|0,x_{1}\rangle$ transforming as $\Gamma^{(5)}$ is
\begin{eqnarray}
|X_{5}^{(2)}\rangle & = &- \frac{1}{3\sqrt{3}}[ (1-e^{-2\pi i/3})|u_{0},u_{1}\rangle \nonumber \\
& & + (1-e^{2\pi i/3})|u_{0},u_{2}\rangle \nonumber \\
& & + \frac{1}{3}(|u_{1},u_{0}\rangle + |u_{2},u_{0}\rangle ) .
\end{eqnarray}
For both $|X_{5}^{(1)}\rangle$ and $|X_{5}^{(2)}\rangle$ we find $\|X_{5\alpha}\|^{2} =  \|X_{5\beta}\|^{2} =2/9$ and $z=-1/(3\sqrt{3})$.  Putting these together, we find the that eigenvalues of $A$ in the $\Gamma^{(5)}$ subspace are $(4/3)(2\pm \sqrt{3})$ each of which is two-fold degenerate.

\end{document}